\documentclass[aps,prc,twoside,twocolumn,nofootinbib,showpacs,showkeys]{revtex4}
\usepackage{bm,epsfig}

\begin{document}

\title[]{Vector Meson Photoproduction Processes Near Threshold}

\author{Yongseok \surname{Oh}}
\email{yoh@phya.yonsei.ac.kr}
\affiliation{Institute of Physics and Applied Physics,
Yonsei University, Seoul 120-749, Korea}
\received{\today}

\begin{abstract}
We discuss the photoproduction processes of light vector mesons ($\rho$,
$\omega$, and $\phi$) from the nucleon near threshold.
We first develop a simple model based on meson exchanges which is modified
by the nucleon pole terms.
We then extend this model to study other physically interesting topics.
As examples, we discuss the missing nucleon resonances problem in
$\omega$ photoproduction and the direct $\phi NN$ coupling constant in
$\phi$ photoproduction.
The calculated cross sections are compared with the recent experimental data.
Various spin observables are discussed, which may be measured at current
photon/electron facilities such as TJNAF and SPring-8 of RCNP.
Precise measurements of such quantities would provide very useful information
to understand the production mechanism.
\end{abstract}


\maketitle

\section{Introduction}

At high energies and low momentum transfers, the exclusive
electromagnetic production of vector mesons has been explained
successfully by the (soft) Pomeron exchange model \cite{DL84,LM95,PL97}.
(For the failure of the soft Pomeron model at large $t$,
see, for example, Refs. \cite{DL98-DL00,KMOV00,OKMV00}.)
However, at low energies near threshold, it is well known that the
meson exchange mechanisms or the exchange of the secondary Regge
trajectories become important or even dominant, of which examples are
the $\pi$ exchange in $\omega$ production and the $\sigma$ exchange in
$\rho$ production \cite{FS96}.
Furthermore, the nucleon exchange in $s$- and $u$-channels is known to be
important in the large momentum transfer region at low energy, although its
contribution becomes suppressed rapidly as the total energy increases.

The process of vector meson photoproduction, such as $\rho$, $\omega$,
and $\phi$, attracts recent interests both theoretically and
experimentally.
This is because the vector meson production processes can give a very
useful or unique probe to investigate several hadron physics problems.
For example, the study of vector meson photoproduction is expected to
be useful to resolve the so-called ``missing resonances'' problem
\cite{CR00,Burk02}.
The constituent quark models predict a much richer nucleon excitation
spectrum than what has been observed so far in the pion-nucleon channel
\cite{IK77-78-79,KI80}.
This has been attributed to the possibility that a lot of the predicted
nucleon resonances ($N^*$) could couple weakly to the $\pi N$ channel.
Therefore it is necessary to search for the nucleon excitations in
other reactions to resolve the missing resonances problem.
Electromagnetic production of vector mesons is one of such reactions and
is being investigated actively with kaon photoproduction \cite{MB99}.
Experimentally, data of vector meson photoproduction in the resonance
region are now being rapidly accumulated at ELSA-SAPHIR of Bonn
\cite{Klein96-98}, Thomas Jefferson National Accelerator Facility
(TJNAF) \cite{CLAS00,CLAS01-CLAS01b-CLAS02}, GRAAL of Grenoble \cite{graal},
LEPS of SPring-8 \cite{leps}, and etc.
Theoretically, there are some recent progress \cite{ZLB98,Zhao01,OTL01} in
this direction.
In this paper, we discuss the role of the nucleon resonances in
$\omega$ photoproduction in the resonance region.

Another interesting topic which will be discussed here is the direct
coupling of the $\phi$ meson to the nucleon.
The electromagnetic production of the $\phi$ from the nucleon has
been suggested as a probe to study the hidden strangeness of
the nucleon \cite{HKPW91,TOY94-97,TYO97,TOYM98,OTYM99-01}.
This is because the $\phi$ is nearly pure $s\bar{s}$ state so that its
direct coupling to the nucleon is suppressed by the OZI rule.
However, if there exists a non-vanishing $s\bar s$ sea quark component
in the nucleon, the strange sea quarks can contribute to the $\phi$
production via the OZI evasion processes.
Investigation of such processes can then be expected to shed light on
the strangeness content of the nucleon, if any \cite{EKKS00}.
In this work, we investigate the direct $\phi NN$ coupling with
polarization observables in $\phi$ photoproduction.

This paper is organized as follows. In Sec. II, we develop models for
vector meson photoproduction based on the Pomeron exchange, one-meson
exchanges, and the nucleon exchange in $s$- and $u$-channels.
By extending this simple model, in Sec. III, we study the nucleon
resonances in $\omega$ photoproduction.
Section IV is devoted to the direct $\phi NN$ coupling study in $\phi$
photoproduction.
Section V contains summary and discussion.

\section{Models for vector meson photoproduction}

\begin{figure}
\centering
\epsfig{file=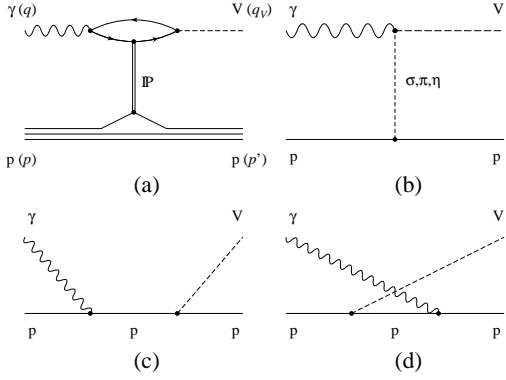,height=2.0in}
\caption{Three mechanisms for vector meson $(V = \rho, \omega, \phi$)
photoproduction: (a) Pomeron, (b) one-meson exchange,
(c) and (d) $s$- and $u$-channel intermediate nucleon diagrams.}
\label{fig:vm}
\end{figure}

In this Section, we briefly discuss a simple model for vector meson
photoproduction \cite{OTL00}.
In order to confront the forthcoming data to extract the nucleon
resonance parameters, it is crucial to understand the non-resonant
(background) production processes which could
interfere strongly with the resonant production amplitudes.
As a step in this direction, we have investigated the $\rho$, $\omega$, and
$\phi$ photoproductions at low energies based on a model consisting of
three mechanisms as shown in Fig.~\ref{fig:vm}: Pomeron exchange,
one-meson ($\pi$, $\eta$, $\sigma$) exchange, and the mechanism involving an
intermediate nucleon in $s$- and $u$-channel
(called nucleon exchange from now on).

\subsection{Pomeron exchange}

We first consider the Pomeron exchange depicted in Fig.~\ref{fig:vm}(a).
In this process, the incoming photon first converts into a $q\bar{q}$
pair, which interacts with the nucleon by the Pomeron exchange before
forming the outgoing vector meson.
The quark-Pomeron vertex is obtained by the Pomeron-photon analogy
\cite{DL84}, which treats the Pomeron as a $C=+1$ isoscalar photon, as
suggested by a study of nonperturbative two-gluon exchanges \cite{LN87}.
We then have
\begin{eqnarray}
T_{fi}^P &=&
i 12 \sqrt{4\pi\alpha_{\rm em}} \beta_u G_P (s,t)
F_1(t) \frac{m_V^2 \beta_f}{f_V}
\nonumber \\ && \mbox{} \times
\frac{1}{m_V^2 - t}
\left( \frac{2\mu_0^2}{2 \mu_0^2 + m_V^2 - t} \right) \varepsilon^*_\mu(V) \varepsilon_\nu(\gamma)
\nonumber \\ && \mbox{} \times
\bar{u}_{m'} (p') \bigl\{
\not \! q \, g^{\mu\nu} - q^\mu \gamma^\nu
\bigr\} u_m(p),
\label{Tpom}
\end{eqnarray}
where $\alpha_{\rm em} = e^2/4\pi$, $m$ and $m'$ are the
spin projections of the initial and final nucleons, respectively
\cite{DL84,LM95,PL97,TOYM98}.
Here we denote the four-momenta of the initial nucleon, final nucleon,
incoming photon, and outgoing vector meson by $p$, $p'$, $q$, and
$q^{}_V$, respectively.
$\varepsilon_\mu^{}(V)$ and $\varepsilon_\nu^{}(\gamma)$ are the
polarization vectors of the vector meson and the photon, respectively.
The Mandelstam variables are $s = W^2 = (p+q)^2$, $t = (p-p')^2$,
$u = (p-q_V^{})^2$.
The vector meson mass is represented by and $m_V^{}$ and $F_1$ is the
isoscalar electromagnetic form factor of the nucleon.
The Pomeron-exchange is described by the Regge form:
\begin{equation}
G_P (s,t) = \left( \frac{s}{s_0^{}} \right)^{\alpha_P^{} (t) - 1}
\exp\left\{ - \frac{i\pi}{2} [ \alpha_P^{} (t) - 1 ] \right\}.
\label{pomG}
\end{equation}
The Pomeron trajectory is taken to be the usual form
$\alpha_P^{} (t) = 1.08 + \alpha'_P t$ with $\alpha'_P = 1/s_0 = 0.25$
GeV$^{-2}$.
In Eq. (\ref{pomG}), $f_V$ is the vector meson decay constant.
The parameters are chosen to
reproduce the total cross section data at high energies $E_\gamma \ge 10$
GeV where the total cross section of vector meson photoproductions are
completely dominated by the Pomeron-exchange.

\subsection{Meson and nucleon exchanges}

For the one-meson exchange ($t$-channel) diagram of Fig.~\ref{fig:vm}(b),
we consider scalar and pseudoscalar meson exchanges.
The vector meson exchange is not allowed in this process and the possible
exchange of axial vector mesons \cite{KMOV00} is suppressed at low
energies mainly because of their heavy masses and small coupling constants.
We also include the nucleon exchanges in $s$- and $u$-channel as in
Fig.~\ref{fig:vm}(c,d).
Those production amplitudes can be calculated from the following
effective Lagrangian:
\begin{equation}
\mathcal{L} = \mathcal{L}_{V\gamma\varphi} + \mathcal{L}_{\varphi NN}
+ \mathcal{L}_\sigma + \mathcal{L}_{\gamma NN} + \mathcal{L}_{VNN},
\label{lagrangian}
\end{equation}
where
\begin{eqnarray}
{\cal L}_{V\gamma\varphi} &=&
\frac{eg_{V\gamma\varphi}}{2M_V^{}}
\varepsilon^{\mu\nu\alpha\beta} \partial_\mu V_\nu^{} \partial_\alpha
A_\beta \varphi,
\end{eqnarray}
where $\varphi = (\pi, \eta)$ and
\begin{eqnarray}
{\cal L}_{\varphi NN} &=& \frac{g_{\pi NN}}{2M_N} \bar\psi \gamma^\mu
\gamma_5 \partial_\mu \pi \psi +
\frac{g_{\eta NN}}{2M_N} \bar\psi \gamma^\mu \gamma_5 \psi \partial_\mu
\eta,
\nonumber \\
{\cal L}_\sigma &=& g_{\sigma NN} \bar\psi \sigma \psi
\nonumber \\ && \mbox{} +
\frac{eg_{\rho\gamma\sigma}}{2M_\rho} \,\mbox{Tr}\, [
\tau^3 \partial_\mu \rho_\nu \left(
\partial^\mu A^\nu - \partial^\nu A^\mu \right) \sigma],
\nonumber \\
{\cal L}_{\gamma NN} &=& -e \bar\psi \left( \gamma_\mu \frac{1+\tau_3}{2}
A^\mu - \frac{\kappa_N}{2M_N} \sigma^{\mu\nu} \partial_\nu A_\mu \right)
\psi,
\nonumber \\
{\cal L}_{VNN} &=& -\frac{g_{\rho NN}}{2} \bar\psi \left( \gamma_\mu
\rho^\mu - \frac{\kappa_\rho}{2M_N} \sigma^{\mu\nu} \partial_\nu
 \rho_\mu \right) \psi
\nonumber \\ && \mbox{}
- g_{\omega NN} \bar\psi \left( \gamma_\mu
\omega^\mu - \frac{\kappa_\omega}{2M_N} \sigma^{\mu\nu} \partial_\nu
 \omega_\mu \right) \psi,
\label{lags}
\end{eqnarray}
with the photon field $A_\mu$, where $\pi$,
$\eta$, $\rho_\mu$
($ = \bm{\tau} \cdot \bm{\rho}_\mu$), and $\omega_\mu$ are the $\pi^0$,
eta, rho, and omega meson fields, respectively.


The coupling constants $g^{}_{V\gamma\pi}$ and $g^{}_{V\gamma\eta}$
are obtained from the experimental
partial widths \cite{PDG98} of the vector meson radiative decays
$V \rightarrow \gamma \varphi$.
We use $g^2_{\pi NN} / 4\pi = 14.0$ and the SU(3) relation to
obtain $g_{\eta NN} / g_{\pi NN} \simeq 0.35$.
To account for the effects due to the finite hadron size at each vertex,
the resulting Feynman amplitudes are regularized by the following form
factors,
\begin{equation}
F_{M NN} = \frac{\Lambda_M^2 - M_M^2}
                {\Lambda_M^2 - t }, \qquad
F_{V\gamma M} = \frac{\Lambda_{V\gamma M}^2 - M_M^2}
{\Lambda_{V\gamma M}^2 - t },
\end{equation}
where ($\Lambda_\pi = 0.7$, $\Lambda_{V\gamma\pi} = 0.77$) and
($\Lambda_\eta = 1.0$, $\Lambda_{V\gamma\eta} = 0.9$) in GeV unit
\cite{TLTS99}.

The scalar ($\sigma$) meson exchange was introduced
to describe the $\rho$ photoproduction \cite{FS96}.%
\footnote{One can describe the $\rho$ photoproduction data by the
exchange of $f_2$ trajectory instead of $\sigma$ exchange
\cite{LM95,DL92}.}
This can be considered as an effective way to account for the two-$\pi$
exchange in $\rho$ production, which is expected to be significant because
of the large branching ratio of $\rho \to \pi^+ \pi^- \gamma$ decay.
We use the parameters of Ref. \cite{FS96} and
$g_{V\gamma\sigma}^{} \simeq 3.0$ to reproduce the total cross sections
of $\rho$ photoproduction near threshold.
We do not consider the $\sigma$ exchange in $\omega$ and $\phi$
photoproductions, since the radiative decays of these two vector mesons
into $\pi^+\pi^-$ are much weaker.
This could be understood by considering the current-field identity
\cite{FS96}. (See also Ref. \cite{TLTS99}.)

The nucleon exchange terms, Fig.~\ref{fig:vm}(c,d), are obtained from
the Lagrangian $\mathcal{L}_{\gamma NN}$ and $\mathcal{L}_{VNN}$, whose
coupling constants are determined from the studies of $\pi N$
scattering and pion photoproduction \cite{SL96}.
In this study, we do not consider the nucleon exchange in $\phi$
photoproduction by assuming $g^{}_{\phi NN} \approx 0$ due to the OZI rule.
This topic will be discussed in Sec. IV.
The form factor $F_V^{}$ for $VNN$ and $\gamma NN$ vertices is assumed to
be of the form given by Ref. \cite{PJ91},
\begin{equation}
F_V (r) =
\frac{\Lambda_{VNN}^4}{\Lambda_{VNN}^4 + (r - M_N^2)^2},
\end{equation}
where $r=(s,u)$ with $\Lambda_{VNN} \simeq 0.8$ GeV.
The gauge invariance is restored by projecting out the gauge
non-invariant part.

\subsection{Results}

With the production amplitudes given above, we have explored the extent
to which the existing data of vector meson photoproductions can be
described by the non-resonant (background) mechanisms.

\begin{figure}[t]
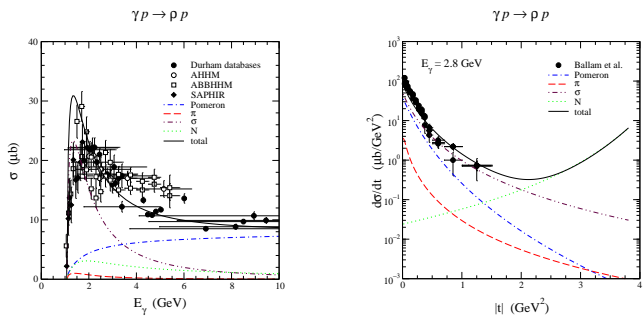

\centering
\epsfig{file=fig2a.eps,height=1.45in, angle=-90} \qquad\quad
\epsfig{file=fig2b.eps,height=1.45in, angle=-90}
\caption{(Left panel) Total cross section of $\rho$ photoproduction.
The experimental data are from Refs. \cite{Durham,AHHM76,ABBH68,Klein96-98}.
(Right panel) Differential cross sections of $\rho$ photoproduction at
$E_\gamma = 2.8$ GeV.
The data are from Refs.~\cite{BCGG72,BCEK73}.}
\label{fig:rho_cs}
\end{figure}


The $\rho$ photoproduction cross sections are calculated from the
amplitudes due to Pomeron, $\pi$, $\sigma$ exchanges and the
nucleon exchange.
In the left panel of Fig.~\ref{fig:rho_cs} we show that the calculated
total cross sections (solid curve) agree to a very large extent with the
data up to $E_\gamma = 10$ GeV.
The important role of the $\sigma$ exchange (dot-dot-dashed curve)
near threshold is also shown there.
The dynamical content of the model can be better seen by investigating
angular distributions.
For example, given in the right panel of Fig.~\ref{fig:rho_cs} is the
angular distributions of the differential cross sections at $E_\gamma =
2.8$ GeV.
We see that the contribution from the nucleon exchange (dotted curves)
becomes dominant at large scattering angles.
Therefore, precise measurements of the differential cross sections in
the backward scattering region will shed light on the role of the
intermediate nucleon states.

\begin{figure}[t]
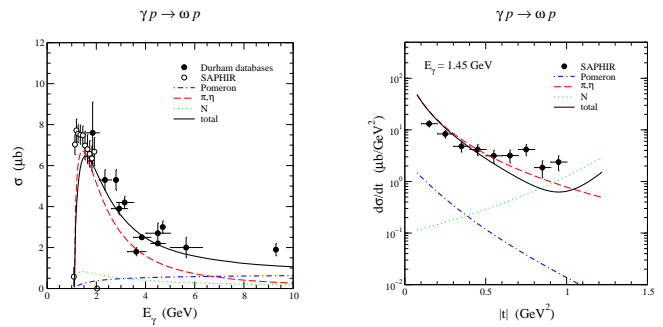

\centering
\epsfig{file=fig3a.eps,height=1.5in, angle=-90} \hfill
\epsfig{file=fig3b.eps,height=1.5in, angle=-90} \hfill
\caption{Total and differential cross sections of $\omega$ photoproduction.
The experimental data are from Refs.
\cite{Durham,Klein96-98,BCGG72,BCEK73}.}
\label{fig:omega_cs}
\end{figure}

The $\omega$ photoproduction cross sections are calculated from the
Pomeron, $\pi$ and $\eta$ exchanges, and the nucleon term.
The predicted total and differential cross sections are compared with
the data in Fig.~\ref{fig:omega_cs}.
We see that the one-pion exchange (long dashed curves) dominates $\omega$
photoproduction up to $E_\gamma \approx 6$ GeV.
The predicted total cross sections somewhat underestimate the recent
SAPHIR data in the $E_\gamma \le 2$ GeV region where the mechanisms
involving the excitation of nucleon resonances are expected to play
some roles.
Similar to the case of $\rho$ photoproduction, the nucleon-term
dominates at large $|t|$.

\begin{figure}[t]
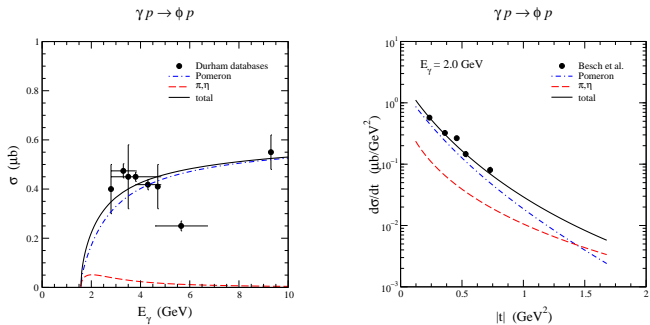

\centering
\epsfig{file=fig4a.eps,height=1.5in, angle=-90} \hfill
\epsfig{file=fig4b.eps,height=1.5in, angle=-90} \hfill
\caption{Total and differential cross sections of $\phi$ photoproduction.
The experimental data are from Refs. \cite{Durham,BHKK74}.}
\label{fig:phi_cs}
\end{figure}

For $\phi$ photoproduction, we consider the Pomeron and pseudoscalar
meson ($\pi$ and $\eta$) exchanges only.
We refer the calculations on the scalar meson exchanges and direct
$\phi$ radiation arising from the non-vanishing $\phi NN$ coupling
to Ref. \cite{TLTS99} and the effects of non-vanishing
strangeness of the nucleon to Refs.~\cite{TOY94-97,TOYM98}.
In Fig.~\ref{fig:phi_cs} we show that the predicted total and
differential cross sections agree well with the very limited data.
Contrary to the cases of $\rho$ and $\omega$ photoproductions, the
Pomeron exchange (dash-dotted curves) gives the major contribution to
the {\em total} cross section even at energies near threshold.
This is mainly due to the fact that the coupling of the $\phi$ to the
nucleon is suppressed by the OZI rule.
The contributions from $\pi$ and $\eta$ exchanges are also found to be
small.

\section{Nucleon resonances in $\bm{\omega}$ photoproduction}

The role of the nucleon excitations in vector meson photoproduction was
studied recently by Zhao {\em et al.\/} \cite{ZLB98,ZDGS99}
using an effective Lagrangian method within the
$\mbox{SU}(6) \times \mbox{O}(3)$ constituent quark model.
In this work, we are motivated by the predictions by Capstick and Roberts
\cite{Caps92,CR94}.
They started with a constituent quark model which accounts for the
configuration mixing due to the residual quark-quark interactions
\cite{GI85-CI86}.
The predicted baryon wave functions are considerably different from those
of the $\mbox{SU}(6) \times \mbox{O}(3)$ model employed by Zhao
{\em et al.\/} in Refs. \cite{ZLB98,ZDGS99}.
The second feature of the predictions from Refs. \cite{Caps92,CR94} is
that the meson decays are calculated from the correlated wave functions
by using the ${}^3P_0$ model.
Thus it would be interesting to see how these predictions differ from
those of Refs. \cite{ZLB98,ZDGS99} and can be tested against the
data of vector meson photoproduction.
We will focus on $\omega$ photoproduction in this work, simply
because its non-resonant reaction mechanisms are much better understood.

In order to estimate the nucleon resonance contributions we make use
of the quark model predictions on the resonance photo-excitation
$\gamma N \to N^*$ and the resonance decay $N^* \to \omega N$ reported
in Refs. \cite{Caps92,CR94} using a relativised quark model.
We calculate the $s$-channel diagrams only since the crossed, $u$-channel,
$N^*$ amplitude cannot be calculated from the informations available in
Refs.~\cite{Caps92,CR94}.
In this study, we consider 12 positive parity and 10 negative parity
nucleon resonances up to spin-$9/2$.
Most of them are ``missing'' so far.
The resonance parameters taken from Refs. \cite{Caps92,CR94} can
be found in Ref. \cite{OTL01} with the other cutoff parameters.
Here we should also mention that we can not account for the
resonances with its predicted masses less than the $\omega N$ threshold,
since their decay vertex functions with an off-shell momentum are
not available. (See Ref.~\cite{TL02} for the contributions from the
$N^*$'s below the threshold.)

\begin{figure}[t]
\centering
\epsfig{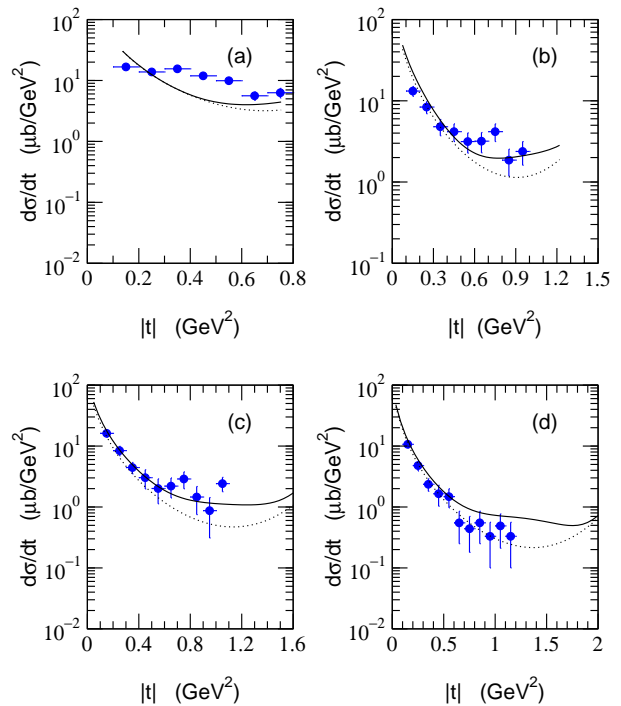}
\caption{
Differential cross sections for the $\gamma p\to p\omega$ reaction
as a function of $|t|$ at different energies: $E_\gamma =$ (a) $ 1.23$,
(b) $1.45$, (c) $1.68$, and (d) $1.92$ GeV.
The solid and dotted curves are calculated respectively with and without
including $N^*$ effects.
Data are taken from Ref. \protect\cite{Klein96-98}.}
\label{fig:dsN*}
\end{figure}

The total resonance effects are shown in Fig.~\ref{fig:dsN*}.
The solid curves are from our full calculations, while the dotted curves
are from the calculations without including $N^*$ excitations.
The results shown in Fig.~\ref{fig:dsN*} indicate that it is rather
difficult to test our predictions by considering only the angular
distributions, since the $N^*$'s influence is mainly in the large
scattering angle region where accurate measurements are perhaps still
difficult.
On the other hand, the forward cross sections seem to be dominated
by the well-understood pseudoscalar-meson exchange and Pomeron exchange.
Therefore, one can use this well-controlled background to examine the
$N^*$ contributions by exploiting the interference effects in the spin
observables.
We also found that the contributions from $N\frac32^+ (1910)$ and
$N\frac32^- (1960)$ are the largest at all energies considered in Fig.
\ref{fig:dsN*}.

\begin{figure}[t]
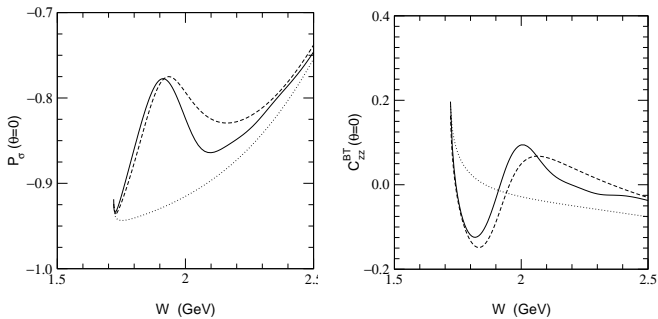

\centering
\epsfig{file=fig6a.eps, width=4.2cm} \hfill
\epsfig{file=fig6b.eps, width=4.2cm}
\caption{
Parity asymmetry $P_\sigma$ at $\theta=0$ as a function of $W$.
The dotted curves are calculated without including $N^*$ effects, the
dashed curves include contributions of $N\frac32^+(1910)$ and
$N\frac32^-(1960)$ only, and the solid curves are calculated with all
$N^*$'s considered in this calculation.}
\label{fig:Psigma}
\end{figure}

To further facilitate the experimental tests of our predictions, we have
investigated spin observables.
We have identified two polarization observables which are sensitive
to the $N^*$ contributions at forward angles, where precise measurements
might be more favorable because the cross sections are peaked at small
$|t|$.
Shown in Fig. \ref{fig:Psigma} are the parity asymmetry $P_\sigma$ and
the beam-target double asymmetry $C_{zz}^{\rm BT}$ at $\theta = 0$,
where $\theta$ is the scattering angle in the center of mass frame.
One can find the sensitivity of those asymmetries on the nucleon
resonances.
Therefore experimental test of our predictions will be a useful step
toward resolving the so-called ``missing resonance problem'' or
distinguishing various QCD-inspired models for hadron.

\section{$\bm{\lowercase{g}_{\phi NN}}$ in $\phi$ photoproduction}

If we assume the ideal mixing of the $\phi$ with the $\omega$, the
direct coupling of the nucleon and the $\phi$ should be suppressed by
the OZI rule.
In nature, the deviation from the ideal mixing is very small and the
direct $\phi NN$ coupling should be very small, although not vanish.
In this Section, we discuss the effective $\phi NN$ coupling in $\phi$
photoproduction.
For this purpose, we use the effective Lagrangian for the $\phi NN$
interaction as in Eq.~(\ref{lags}).

Even if we assume the ideal mixing between the $\phi$ and the $\omega$,
non-vanishing effective $\phi NN$ coupling is allowed by the kaon loops
and hyperon excitations.
The estimated $\phi NN$ coupling is very small, $g_{\phi NN}^{} \simeq
-0.24$ \cite{MMSV97}, which is consistent with the OZI rule prediction.
However the analyses on the electromagnetic nucleon form factors
\cite{GHo76,HPSB76,Jaf89} and the baryon-baryon scattering
\cite{NRS75-77-79} favor a large $\phi NN$
vector coupling constant which strongly violates the OZI rule:
$g_{\phi NN}^{} / g_{\omega NN} = -0.3 \sim -0.43$, therefore
$g_{\phi NN}^{} = -2.3 \sim -4.7$ with
$g_{\omega NN}^{} = 7.0 \sim 11.0$ \cite{SL96}.
In our study the nucleon pole terms are responsible to the $\phi$
photoproduction at large $|t|$ and we find that $|g_{\phi NN}^{}| = 3.0$
can reproduce the recent CLAS data \cite{OB01}, which is consistent with
the Regge trajectory study of Ref.~\cite{Lage00}.

\begin{figure}[t]
\centering
\epsfig{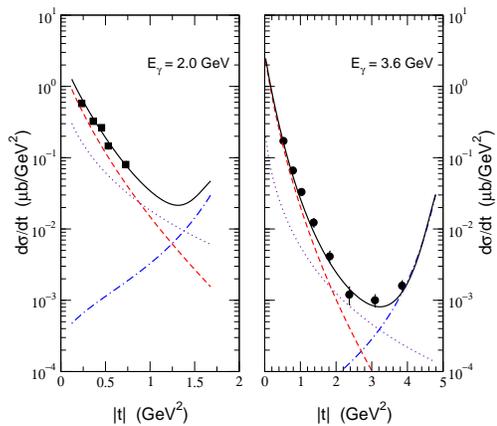}
\caption{Differential cross sections of $\phi$ photoproduction at
$E_\gamma = $ $2.0$ GeV and $3.6$ GeV.
The results are from Pomeron exchange (dashed), pseudoscalar-meson
exchange (dotted), nucleon pole terms with $g_{\phi NN}^{} =
-3.0$ (dot-dashed),
and the full amplitude (solid).
The experimental data are from Ref. \protect\cite{BHKK74} (filled squares)
and Ref. \protect\cite{CLAS00} (filled circles).}
\label{fig:dsdt}
\end{figure}

\begin{figure}[b]
\centering
\epsfig{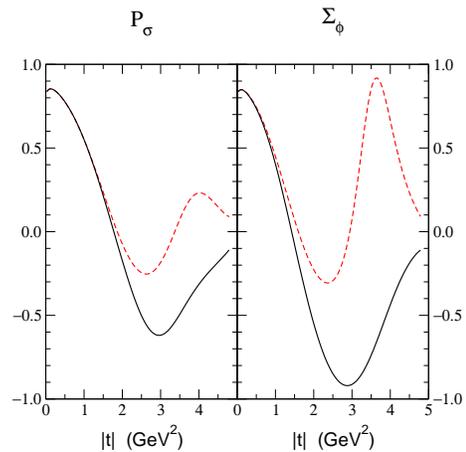}
\caption{Parity asymmetry $P_\sigma$ and the photon polarization
asymmetry $\Sigma_\phi$ in $\phi$ photoproduction at $E_\gamma = 3.6$ GeV
with the full amplitude.
The solid lines are with $g_{\phi NN}^{} = -3.0$ and the dashed lines
with $g_{\phi NN}^{} = +3.0$.}
\label{fig:asym3_6}
\end{figure}

In Fig.~\ref{fig:dsdt}, we compare our predictions with the recent CLAS
data \cite{CLAS00} by assuming $g_{\phi NN}^{}=-3.0$.
The role of the tensor coupling is found to be negligible if
$\kappa_\phi < 1$.
We also found that vector meson density matrix and some asymmetries are
sensitive to the phase of the $\phi NN$ coupling constant.
As an example, we present our prediction of the parity asymmetry
$P_\sigma$ and the photon polarization asymmetry $\Sigma_\phi$ in
Fig.~\ref{fig:asym3_6}.
The details can be found in Ref.~\cite{OB01}.

\section{Summary and Discussion}

In this paper, we briefly review models to understand the production
mechanism of vector meson photoproduction near threshold.
We first discuss models based on the Pomeron, meson, and nucleon
exchanges.
With those background amplitudes, the role of nucleon resonances is
studied in $\omega$ photoproduction and we found that measuring
polarization asymmetries is very useful to identify the contributions
from the nucleon resonances.
The effective $\phi NN$ coupling is also discussed within $\phi$
photoproduction.
The results show that the recent CLAS experiment could be understand by
a fairly large $\phi NN$ coupling constant.
Other physical quantities are suggested to confirm and test the model
predictions.

Finally we have to mention that the extraction of $N^*$ parameters from
experimental data depends strongly on the accuracy of the nonresonant
amplitudes.
Therefore, it is legitimate to develop a dynamical model for the
coupled-channel effects, which should satisfy the unitarity condition
\cite{OL02,PM02d}.
The importance of the final state interaction to extract the direct
$\phi NN$ coupling should also be studied.
Together with those theoretical studies, it is very crucial to test the
models by experiments.
Therefore, experimental research at, for example, TJNAF and SPring-8
would be very important to understand the production mechanisms of
the electromagnetic production of vector mesons and the hadron structure.

\bigskip

I have benefited from discussions and collaboration with H.~Bhang,
N.~I. Kochelev, T.-S.~H. Lee, D.-P. Min, T.~Morii, A.~I. Titov, and
S.~N. Yang.
I am also grateful to V.~D. Burkert, M.~Fujiwara, Hungchong Kim,
Su~Houng Lee, T.~Nakano, E.~Smith, and V. Vento for fruitful discussions.
This work was supported by the Brain Korea 21 project of Korean Ministry
of Education and the International Collaboration Program of
KOSEF under Grant No. 20006-111-01-2.



\begin{thebibliography}{10}

\bibitem{DL84}
A.~Donnachie and P.~V. Landshoff,
  Nucl. Phys. B {\bf 244}, 322 (1984).

\bibitem{LM95}
J.-M. Laget and R.~Mendez-Galain,
  Nucl. Phys. A {\bf 581}, 397 (1995).

\bibitem{PL97}
M.~A. Pichowsky and T.-S.~H. Lee,
  Phys. Rev. D {\bf 56}, 1644 (1997).

\bibitem{DL98-DL00}
A.~Donnachie and P.~V. Landshoff,
  Phys. Lett. B {\bf 437}, 408 (1998);
  {\bf 478}, 146 (2000).

\bibitem{KMOV00}
N.~I. Kochelev, D.-P. Min, Y.~Oh, V.~Vento, and A.~V. Vinnikov,
  Phys. Rev. D {\bf 61}, 094008 (2000);
  Nucl. Phys. B (Proc. Suppl.) {\bf 99}, 24 (2001).

\bibitem{OKMV00}
Y.~Oh, N.~I. Kochelev, D.-P. Min, V.~Vento, and A.~V. Vinnikov,
  Phys. Rev. D {\bf 62}, 017504 (2000).

\bibitem{FS96}
B.~Friman and M.~Soyeur,
  Nucl. Phys. A {\bf 600}, 477 (1996).

\bibitem{CR00}
S.~Capstick and W.~Roberts,
  Prog. Part. Nucl. Phys. {\bf 45}, S241 (2000).

\bibitem{Burk02}
V.~D. Burkert,
  hep-ph/0207149;
  hep-ph/0210321.

\bibitem{IK77-78-79}
N.~Isgur and G.~Karl,
 Phys. Lett. {\bf 72B}, 109 (1977);
 Phys. Rev. D {\bf 18}, 4187 (1978);
 {\bf 19}, 2653 (1979), {\bf 23}, 817(E) (1981).

\bibitem{KI80}
R.~Koniuk and N.~Isgur,
  Phys. Rev. D {\bf 21}, 1868 (1980).

\bibitem{MB99}
T.~Mart and C.~Bennhold,
  Phys. Rev. C {\bf 61}, 012201 (1999);
J.~K. Ahn, in these proceedings.

\bibitem{Klein96-98}
F.~J. Klein, Ph.D. thesis, Bonn Univ. (1996);
 $\pi$N Newslett. {\bf 14}, 141 (1998).

\bibitem{CLAS00}
\mbox{CLAS Collaboration,} E.~Anciant {\em et~al.\/},
  Phys. Rev. Lett. {\bf 85}, 4682 (2000).

\bibitem{CLAS01-CLAS01b-CLAS02}
\mbox{CLAS Collaboration,} K.~Lukashin {\em et~al.\/},
  Phys. Rev. C {\bf 63}, 065205 (2001);
M.~Battaglieri {\em et~al.\/},
  Phys. Rev. Lett. {\bf 87}, 172002 (2001);
M.~Battaglieri {\em et~al.\/},
  hep-ex/0210023.

\bibitem{graal}
J.~Ajaka {\em et~al.\/},
  Talk at SPIN 2000,
  AIP Conf. Proc. No. {\bf 570} (2001) p. 198.

\bibitem{leps}
T.~Nakano,
  Talk at SPIN 2000,
  AIP Conf. Proc. No. {\bf 570} (2001) p. 189;
 in these proceedings.

\bibitem{ZLB98}
Q.~Zhao, Z.~Li, and C.~Bennhold,
  Phys. Lett. B {\bf 436}, 42 (1998).
  Phys. Rev. C {\bf 58}, 2393 (1998).

\bibitem{Zhao01}
Q.~Zhao,
  Phys. Rev. C {\bf 63}, 025203 (2001).

\bibitem{OTL01}
Y.~Oh, A.~I. Titov, and T.-S.~H. Lee,
  Phys. Rev. C {\bf 63}, 025201 (2001).

\bibitem{HKPW91}
E.~M. Henley, G.~Krein, S.~J. Pollock, and A.~G. Williams,
  Phys. Lett. B {\bf 269}, 31 (1991);
E.~M. Henley, G.~Krein, and A.~G. Williams,
  {\em ibid.\/} {\bf 281}, 178 (1992).

\bibitem{TOY94-97}
A.~I. Titov, Y.~Oh, and S.~N. Yang,
  Chin. J. Phys. (Taipei) {\bf 32}, 1351 (1994);
  Phys. Rev. Lett. {\bf 79}, 1634 (1997).

\bibitem{TYO97}
A.~I. Titov, S.~N. Yang, and Y.~Oh,
  Nucl. Phys. A {\bf 618}, 259 (1997).

\bibitem{TOYM98}
A.~I. Titov, Y.~Oh, S.~N. Yang, and T.~Morii,
  Phys. Rev. C {\bf 58}, 2429 (1998).

\bibitem{OTYM99-01}
Y.~Oh, A.~I. Titov, S.~N. Yang, and T.~Morii,
  Phys. Lett. B {\bf 462}, 23 (1999);
  Nucl. Phys. A {\bf 684}, 354 (2001).

\bibitem{EKKS00}
J.~Ellis, M.~Karliner, D.~E. Kharzeev, and M.~G. Sapozhnikov,
  Nucl. Phys. A {\bf 673}, 256 (2000);
J.~Ellis,
  Nucl. Phys. A {\bf 684}, 53 (2001).

\bibitem{OTL00}
Y.~Oh, A.~I. Titov, and T.-S.~H. Lee,
  Talk at NSTAR 2000 Workshop, nucl-th/0004055.

\bibitem{LN87}
P.~V. Landshoff and O.~Nachtmann,
  Z. Phys. C {\bf 35}, 405 (1987).

\bibitem{PDG98}
\mbox{Particle Data Group,} C.~Caso {\em et~al.\/},
  Eur. Phys. J. C {\bf 3}, 1 (1998).

\bibitem{TLTS99}
A.~I. Titov, T.-S.~H. Lee, H.~Toki, and O.~Streltsova,
  Phys. Rev. C {\bf 60}, 035205 (1999).

\bibitem{DL92}
A.~Donnachie and P.~V. Landshoff,
  Phys. Lett. B {\bf 296}, 227 (1992).

\bibitem{SL96}
T.~Sato and T.-S.~H. Lee,
  Phys. Rev. C {\bf 54}, 2660 (1996).

\bibitem{PJ91}
B.~C. Pearce and B.~K. Jennings,
  Nucl. Phys. A {\bf 528}, 655 (1991).

\bibitem{Durham}
\mbox{The Durham RAL Databases},
  http://durpdg.dur.ac.uk/HEPDATA/REAC.

\bibitem{AHHM76}
W.~Struczinski {\em et~al.\/},
  Nucl. Phys. B {\bf 108}, 45 (1976).

\bibitem{ABBH68}
R.~Erbe {\em et~al.\/},
  Phys. Rev. {\bf 175}, 1669 (1968).

\bibitem{BCGG72}
J.~Ballam {\em et~al.\/},
  Phys. Rev. D {\bf 5}, 545 (1972).

\bibitem{BCEK73}
J.~Ballam {\em et~al.\/},
  Phys. Rev. D {\bf 7}, 3150 (1973).

\bibitem{BHKK74}
H.~J. Besch {\em et~al.\/},
  Nucl. Phys. B {\bf 70}, 257 (1974).

\bibitem{ZDGS99}
Q.~Zhao, J.-P. Didelez, M.~Guidal, and B.~Saghai,
  Nucl. Phys. A {\bf 660}, 323 (1999).

\bibitem{Caps92}
S.~Capstick,
  Phys. Rev. D {\bf 46}, 2864 (1992).

\bibitem{CR94}
S.~Capstick and W.~Roberts,
  Phys. Rev. D {\bf 49}, 4570 (1994).

\bibitem{GI85-CI86}
S.~Godfrey and N.~Isgur,
  Phys. Rev. D {\bf 32}, 189 (1985);
S.~Capstick and N.~Isgur,
  {\em ibid.\/} {\bf 34}, 2809 (1986).

\bibitem{TL02}
A.~I. Titov and T.-S.~H. Lee,
  Phys. Rev. C {\bf 66}, 015204 (2002).

\bibitem{MMSV97}
U.-G. Mei{\ss}ner, V.~Mull, J.~Speth, and J.~W. Van~Orden,
  Phys. Lett. B {\bf 408}, 381 (1997).

\bibitem{GHo76}
H.~Genz and G.~H{\"o}hler,
  Phys. Lett. {\bf 61B}, 389 (1976).

\bibitem{HPSB76}
G.~H{\"o}hler {\em et~al.\/},
  Nucl. Phys. B {\bf 114}, 505 (1976).

\bibitem{Jaf89}
R.~L. Jaffe,
  Phys. Lett. B {\bf 229}, 275 (1989).

\bibitem{NRS75-77-79}
M.~M. Nagels, T.~A. Rijken, and J.~J. de~Swart,
  Phys. Rev. D {\bf 12}, 744 (1975);
  {\bf 15}, 2547 (1977);
  {\bf 20}, 1633 (1979).

\bibitem{OB01}
Y.~Oh and H.~C. Bhang,
  Phys. Rev. C {\bf 64}, 055207 (2001).

\bibitem{Lage00}
J.~M. Laget,
  Phys. Lett. B {\bf 489}, 313 (2000).

\bibitem{OL02}
Y.~Oh and T.-S.~H. Lee,
  Phys. Rev. C {\bf 66}, 045201 (2002);
  Talk at PANIC 2002 (2002), nucl-th/0211054.

\bibitem{PM02d}
G.~Penner and U.~Mosel,
  nucl-th/0207069.


\end{thebibliography}
\end{document}